\documentclass{cpp2e}
\usepackage{epsfig,latexsym,amssymb}
\newcommand{\be}{\begin{equation}}
\newcommand{\ee}{\end{equation}}
\newcommand{\ba}{\begin{eqnarray}}
\newcommand{\ea}{\end{eqnarray}}

\shorttitle{Interpolation formula}
\shortauthor{A. Esser, R. Redmer, G. R\"opke}
\contvolum{} \contyear{} \contnumber{} \contpages{}

\title{Interpolation formula for the electrical conductivity of  
 nonideal plasmas}

\author{A.~Esser$^{\rm a}$, R.~Redmer$^{b}$, G.~R\"opke$^{b}$} 

\address{
$^{\rm a}$ Humboldt-Universit\"at zu Berlin, Institut f\"ur Physik, 
  D-10117 Berlin, Germany\\
$^{\rm b}$ Universit\"at Rostock, Fachbereich Physik, D-18051
  Rostock, Germany
}

\begin{document}
\setcounter{page}{1}
\makeheadings
\maketitle

\begin{abstract}
  On the basis of a quantum-statistical approach to the electrical
  conductivity of nonideal plasmas we derive analytical results in the
  classical low-density regime, in the degenerate Born limit, and for
  the contribution of the Debye-Onsager relaxation effect. These
  explicit results are used to construct an improved interpolation
  formula of the electrical conductivity valid in a wide range of
  temperature and density which allows to compare with available
  experimental data of nonideal plasmas.
\end{abstract}

\section{Introduction}

Optical and transport properties of plasmas are governed by the mutual
Coulomb interaction and depend strongly on temperature and the
electron and ion density. Experimental efforts for the measurement of
the electrical conductivity up to high densities revealed the
importance of collective excitations and many-body effects such as
Pauli blocking, dynamic screening and self-energy, structure factor,
the Debye-Onsager relaxation effect, formation of bound states
etc.\ \cite{KKER}.  Although different methods have been proposed for
the evaluation of plasma transport properties
\cite{Ichimaru,Ebeling,Lee,Rinker,Tkach}, it still remains a
theoretical challenge to treat these effects by a unified 
quantum-statistical approach. 

On the other hand, in addition to highly sophisticated approaches to
the conductivity accounting for quantum statistical many-particle
effects, for practical use interpolation formulas are of interest
which are applicable in a large region of plasma parameter values.
Such interpolation formulas were developed by several authors
\cite{Ichimaru,Ebeling,Lee,Rinker} in order to evaluate complex
physical situations, e.g., in hydrodynamical simulations for the
generation and expansion of plasmas produced by high-power lasers,
energetic heavy ion beams or magnetic compression (pinches). Based on
rigorous results for special limiting cases and possibly numerical
simulations in the intermediate regions, Pad\'e-type interpolation
formulas have been proven to be highly effective to describe
thermodynamic as well as tranport properties.
 
In this paper we consider a fully ionized plasma consisting of
electrons (mass $m_e$) and singly charged ions (mass $M$), interacting
via the Coulomb potential, at density $n_e=n_i=n$ and temperature $T$
(hydrogen plasma). The dimensionless parameters 
\ba \label{pp}
\Gamma &=& {e^2 \over {4\pi\epsilon_0k_BT}}
 \left({4\pi n \over 3}\right)^{\!1/3} \,,\,
 \Theta = {2m_ek_BT \over \hbar^2}(3 \pi^2 n)^{-2/3}\,,
\ea
describe the ratio between the mean potential energy and the kinetic
energy ($\Gamma$) and denote the degree of degeneracy of the electron
gas ($\Theta$). Usually, plasmas are classified into ideal ($\Gamma
\ll 1$) and nonideal ($\Gamma \ge 1$) as well as degenerate
($\Theta\ll 1$) and nondegenerate ($\Theta\gg 1$) systems.

Using these dimensionless plasma parameters, the construction of
interpolation formulas for the dc conductivity has been performed in
different works. Ichimaru and Tanaka \cite{Ichimaru} considered a
two-component plasma at finite temperatures within a generalized Ziman
formula by taking into account dynamic local field effects in the
dielectric function and the dynamic ion-ion structure factor. The
collision integrals were evaluated in Born approximation to study a
strongly coupled and degenerate plasma with $\Gamma \geq 1$ and
$\Theta \leq 1$. The correct nondegenerate limit was incorporated in a
semiempirical way by adopting a prefactor in correspondence to the
well-known Spitzer formula.

Ebeling {\it et al.}  \cite{Ebeling} and Lee and More \cite{Lee} used
the relaxation time approximation for the evaluation of the collision
integrals both in the low-density (nondegenerate) and high-density
(degenerate) regime. Electron-electron interactions are neglected in
that approximation which is only justified in the degenerate case
because of Pauli blocking. The consideration of electron-electron
scattering is, however, decisive to obtain the right low-density
result, the prefactor of the Spitzer formula. This has been modeled
using a semiempirical ansatz. Furthermore, Rinker
\cite{Rinker} has derived an interpolation formula for the electrical 
conductivity from an improved Ziman formula which are applicable in
the strongly coupled, degenerate domain.

Within a quantum statistical approach, the transport properties within
the generalized linear response theory of Zubarev have been evaluated
\cite{Roe1988} and rigorous analytical results for the electrical
conductivity have been derived which are valid in the low-density
(nondegenerate) and high-density (degenerate) limit
\cite{Roe1988,RR,RRR}. The influence of non-equilibrium two-particle
correlations in lowest order (Debye-Onsager effect) has been
reconsidered recently \cite{ER}. Based on these results and taking
into account other approaches, we will construct an improved
interpolation formula for the electrical conductivity to cover a wide
range of temperatures and densities.

\section{Electrical conductivity}

Using the plasma parameters (\ref{pp}), we can express the electrical 
conductivity as 
\be 
\sigma(n,T) = {(k_B T)^{3/2}(4\pi\epsilon_0)^{2} \over {m_e^{1/2} e^2}}
\sigma^*(\Gamma,\Theta) 
\ee 
with a universal function $\sigma^*(\Gamma,\Theta)$ depending on the
characteristic plasma parameters exclusively. 

Our quantum-statistical approach to the electrical conductivity is
based on a generalized Boltzmann equation which is derived from
linear response theory \cite{Roe1988}. A finite-moment approximation
of single and two-particle distribution functions defines the set of
relevant observables $B_n$, and the electrical conductivity is 
obtained from the ratio
\be
\sigma = -{\beta \over \Omega }{1 \over {|D[B_{n'};B_{n}]|}}
\left| \begin{array}{ll}   0 &  N[B_n] \\
Q[B_n] & D[B_{n'};B_{n}]
\end{array} \right| \,.
\label{cond}
\ee 
The determinants contain equilibrium correlation functions defined by
\ba
N[B_n] &=& (\dot R; B_n) \,,\\
Q[B_n] &=& (\dot R; B_n)+\langle \dot R; \dot B_n \rangle \,,\nonumber\\
D[B_{n'};B_{n}] &=& (B_{n'}; \dot B_{n})+ \langle \dot B_{n'}; 
 \dot B_{n} \rangle \,,\nonumber 
\label{elements}
\ea
which are given by Kubo's scalar product between two operators:
\ba
(A;B) &=& {1 \over \beta}\int_0^{\beta} {\rm d}\tau\, 
 {\rm Tr}[\rho_0 A(-i \hbar \tau)B] \,,\\
\langle A(\epsilon);B \rangle &=& \lim_{\epsilon \rightarrow 0} 
 \int_{-\infty}^0 {\rm d}t\, e^{\epsilon t} (A(t);B) \,.\nonumber
\ea
The two-particle center of mass momentum is defined via
$P=-em_eM/(m_e+M)\dot R$. $\Omega$ is the system volume and
$\beta=1/k_BT$. Furthermore, the time-dependence of the relevant
observables $B_n$ and of the equilibrium statistical operator $\rho_0$
is given by the standard system Hamiltonian of a fully ionized plasma,
see Ref.\ \cite{Roe1988}.  The correlation functions $D[B_{n'};B_{n}]$
are related to thermodynamic Green functions. Thus, efficient diagram
techniques provide a systematic treatment of many-particle effects in
a strongly coupled plasma which has been demonstrated for the single
and two-particle level, see Refs.\ \cite{Roe1988,ER}.

As the relevant observables on the single-particle level are treated
the moments of the electron distribution (adiabatic approximation)
\ba
P_m = \sum_k \hbar k_z \left({\beta \hbar^2 k^2 \over 2 m_e}\right)^{\!m} 
a^\dagger_e (k) a^{\,}_e (k) \,,
\ea
including the total electron momentum ($P_0$) and the ideal energy
current ($P_1$).  A systematic increase in the number of moments
results in a converging expression of the conductivity as it is known
from the Kohler variational principle or the Grad and Chapman-Enskog
method of kinetic theory.

The correlation functions $D[P_n,P_m]$ are given by a sum of ladder
diagrams in the low-density limit. Both the electron-electron and the
electron-ion scattering processes are included, see Ref.\ \cite{RR}.
The scattering integrals have been evaluated in various approximations
which are appropriate for a given density and temperature range.

(i) The correlation functions in the classical limit ($\Gamma^2\Theta
\gg 1$) are related to Boltzmann collision integrals which are given
by transport cross sections. These quantities are treated in the
quasi-classical approximation as relevant for low-energy particles.
Additional quantum corrections are found from a WKB expansion of the
collison integrals.  Using a three-moment approximation for the 
one-particle distribution function \cite{RRT} we find for 
$\Gamma^2\Theta\gg 1$ 
\ba
\sigma^*(\Gamma^2 \Theta\gg 1) &=& 2 a \left( \ln \Gamma^{-3} + 
  0.2943 - {0.523 \over \Gamma^2 \Theta} - {0.259 \over \Gamma^4
  \Theta^2} \right)^{\!-1} .
\ea
The prefactor $a=0.591$ is the limiting value of the moment expansion
and coincides with the Spitzer result \cite{Spitzer}.

(ii) Born limit ($\Gamma^2\Theta\ll 1$): The correlation functions
are equivalent to the Lenard-Balescu collision integral if the
random phase approximation (RPA) is considered for the dielectric
function $\epsilon_{\rm RPA}(q,\omega)$.  The following low-density 
limit is obtained for $\Gamma^2\Theta\ll 1$ \cite{RRMK}:
\be
\sigma^{*}(\Gamma^2 \Theta \ll 1) = 2a \left(\ln {\Theta \over \Gamma}
 + 0.4807 \right)^{\!-1} .
\ee

(iii) The correlation functions in the high-density (liquid-like)
limit are given by Landau collision integrals which treats the
electron-ion scattering in Born approximation. Thus, the corresponding 
electron-ion contribution is
\ba \label{Landau}
D[P_n;P_m] \!&=&\! {\Omega^2 m_e^2 N_i^2 \over 12 \pi^3 \hbar^3}
 \!\int_0^{\infty}\!\!dk \left(-{d f_e(k) \over dk}\right)\! 
 \left({\beta \hbar^2 k^2 \over 2 m_e}\right)^{\!m+n} 
 \!\int_0^{2k}\!\!dq\,q^3 S_{ii}(q)\left|{V_{ei}(q) \over 
 \epsilon(q)}\right|^2 .
\ea
The number of required moments $P_m$ reduces with increasing density;
the Ziman-Faber formula is already obtained from Eq.\,(\ref{Landau})
with the lowest moment $P_0$. It includes the static ionic structure
factor $S_{ii}(q)$, the electron-ion pseudopotential $V_{ei}(q)$, and
the static dielectric function of the electrons $\epsilon(q)$.

The consideration of two-particle correlations described by respective 
moments 
\ba
\delta n^{m,m'}_{cd}(q) &=& \sum_{k,p} f_{c,d}^{m,m'}({\bf k,p,q}) 
  a^\dagger_c(k-\frac{q}{2}) a^\dagger_d(p+\frac{q}{2})
  a_d(p-\frac{q}{2}) a_c(k+\frac{q}{2})  
\ea
leads to a decreased plasma conductivity. Spatial symmetry properties
of the two-particle distribution function in the case of an applied
electric field restricts the moments to the electron-ion contribution 
$\delta n^{m,m'}_{ei}(q)$ \cite{ER}. The evaluation of the
first moment $f^{0,0}_{ei}=1$ in the low-density regime gives
the original Onsager result of the Debye-Onsager relaxation effect 
\cite{Roe1988,ER}.  This effect decreases the conductivity due to an 
incomplete formation of the screening cloud and was first considered 
in the theory of electrolytes \cite{Onsager}. An equivalent 
formulation is the hydrodynamic approximation in kinetic theory and 
assumes a local equilibrium for the distribution functions
\cite{hydro}. As this assumption already fails to describe the 
single-particle properties (Spitzer formula) properly also higher 
two-particle moments have been considered in Ref.\ \cite{ER}.
Corrections to the Onsager result were indeed found for a fully
ionized plasma from the second moments $\delta n^{2,0}_{ei}(q)$ with 
$f^{2,0}_{ei}=k^2$ and a virial expansion of the conductivity 
\be
\sigma^* \approx a(\ln \Gamma^{-3/2}+b+c~\Gamma^{3/2} \ln 
 \Gamma^{-3/2})^{-1}  
\label{sigma2}
\ee
could be derived in the classical low-density and nondegenerate limit.
We find for hydrogen plasmas

\ba \label{sigma3}
\sigma &=& 0.591 {(4\pi \epsilon_0)^2(k_B T)^{3/2} \over {e^2
   m^{1/2}}} \left[ \ln \Gamma^{-3/2} + 1.124 + {1.078 \over
   {\sqrt{6}+\sqrt{3}}} \Gamma^{3/2} \ln\Gamma^{-3/2} \right]^{-1}
   \\ 
&=& 1.530 \times 10^{-2} T^{3/2}\left[ \ln\Gamma^{-3/2} + 1.124
   \,+\, 0.258 \Gamma^{3/2} \ln\Gamma^{-3/2} \right]^{-1} 
   (\Omega {\rm m}{\rm K}^{3/2})^{-1} \,.\nonumber
\ea
The inclusion of the two-particle nonequilibrium correlations 
determines the coefficient $c$ of the term $\Gamma^{3/2}{\rm ln} 
\Gamma^{-3/2}$ in Eq.\ (\ref{sigma2}).

\section{Interpolation formula and results}

An interpolation formula can be constructed on the basis of the
explicit analytical results given above so that reliable results for
the electrical conductivity are obtained easily for a wide range of
density and temperature without the necessity to evaluate the
underlying complicated many-particle methods in full detail.  The
original interpolation formula given in Ref.\ \cite{RRR} included the
low-density ($\Gamma \ll 1$) and Born limit ($\Gamma^2\Theta \ll 1$).
We improve the validity region and the accuracy of that formula in the
present paper by (i) taking into account the corrections due to the
Debye-Onsager relaxation effect and (ii) by a better analytical
structure of the interpolation formula, avoiding unphysical behavior
such as, e.g., poles in the entire ($\Gamma, \Theta$)-plane. 
(iii) Furthermore we incorporate the results of 
Ichimaru and Tanaka in the strongly coupled and 
degenerate limit ($\Theta \le 1, \Gamma \ge 1$) by analyzing their 
parameterized numerical results. Thus the validity range 
of our interpolation formula is extended to a parameter range where the 
influence of the ion-ion structure factor becomes relevant. 

We propose the following interpolation formula 
[$T$ in K, $\sigma$ in ($\Omega$m)$^{-1}$]: 
\ba\label{interneu}
\sigma &=& a_0 T^{3/2}\left(1+{b_1 \over \Theta^{3/2}}\right) 
  \left[\ln(1+A+B)D-C-{b_2 \over b_2+\Gamma\Theta} \right]^{-1} 
\ea
with the functions
\ba
A &=& \Gamma^{-3}{1+a_4/\Gamma^2 \Theta \over 1+a_2/\Gamma^2 
 \Theta+a_3/\Gamma^4 \Theta^2} \left[a_1+c_1 \ln (c_2 \Gamma^{3/2}+1) 
 \right]^2 \,,\\
B &=& {b_3(1+c_3 \Theta) \over \Gamma \Theta(1+c_3 \Theta^{4/5})}
 \,,\nonumber\\ 
C &=& {c_4 \over \ln(1+\Gamma^{-1})+c_5 \Gamma^2 \Theta} 
 \,,\nonumber\\
D &=& {\Gamma^{3}+a_5(1+a_6 \Gamma^{3/2}) \over \Gamma^{3}+a_5}
 \,.\nonumber 
\ea
The parameters $a_i$ are fixed by the low-density virial expansion. In
particular, the corrections from the Debye-Onsager relaxation effect
are included in the function $D$. The coefficients $b_i$ are used 
to adjust the Ichimaru and Tanaka results in the strong degenerate 
limit, and the 
parameters $c_i$ were fitted to numerical data of the correlation 
functions. The explicit set of parameters is given by 
$a_0=0.03064$, $a_1=1.1590$, $a_2=0.698$, $a_3=0.4876$, $a_4=0.1748$, 
$a_5=0.1$, $a_6=0.258$, $b_1=1.95$, $b_2=2.88$, $b_3=3.6$, $c_1=1.5$,
$c_2=6.2$, $c_3=0.3$, $c_4=0.35$, $c_5=0.1$.

\begin{table}[h]
\caption{Electrical conductivity according to the new interpolation 
  formula (\ref{interneu}) compared with the former one given in 
  \protect\cite{RRR}, the Ichimaru-Tanaka fit formula
  \protect\cite{Ichimaru}, and available experimental data for 
  strongly coupled plasmas \protect\cite{Ivanov,Popovic,Radtke}. 
\label{table}}
\begin{center}
\begin{tabular}{lllllrrrr} \hline
 & T & $n_e$ & $\Gamma$ & $\,\,\,\,\Theta$ &
 \multicolumn{4}{c}{$\sigma (10^2\,\Omega^{-1}$m$^{-1})$} \\
 & ($10^3$K) & ($10^{25}\!$/m$^{3})$ & & 
 & exp. & $[2]$ & $[9]$ & $(13)$ \\
\hline
Ar\cite{Ivanov} & 22.2 & 2.8 & 0.368 &~ 56.9 &~ 190  & 200 &225 &220\\
                & 20.3 & 5.5 & 0.505 &~ 33.2 &~ 155  & 203 &231 &224\\
                & 19.3 & 8.1 & 0.604 &~ 24.4 &~ 170  & 209 &239 &232\\
                & 19.0 & 14  & 0.736 &~ 16.7 &~ 255  & 234 &269 &261\\
                & 17.8 & 17  & 0.838 &~ 13.7 &~ 245  & 232 &270 &262\\
Xe\cite{Ivanov} & 30.1 & 25  & 0.564 &~ 17.9 &~ 450  & 442 &458 &453\\
                & 27.0 & 79  & 0.922 &~ 7.47 &~ 740  & 546 &594 &590\\
                & 25.1 & 160 & 1.26  &~ 4.34 &~ 780  & 660 &813 &797\\
                & 22.7 & 200 & 1.50  &~ 3.38 &~ 930  & 694 &933 &900\\
Ne\cite{Ivanov} & 19.8 & 1.1 & 0.303 &~ 94.6 &~ 130  & 148 &173 &169\\
                & 19.6 & 1.9 & 0.367 &~ 65.0 &~ 165  & 160 &187 &181\\
Air\cite{Ivanov}& 11.0 & 0.13& 0.267 &~ 218  &~ 60   & 53  &67  &65 \\
\hline
Ar\cite{Popovic}& 16.4 & 0.06 & 0.128 &~ 551 &~ 83  & 78 &93  &93\\
                &      & 0.1  & 0.165 &~ 385 &~ 79  & 83 &102 &102\\
                &      & 0.13 & 0.18  &~ 324 &~ 76  & 86 &105 &105\\
                &      & 0.15 & 0.19  &~ 291 &~ 64  & 88 &108 &107\\
Xe\cite{Popovic}& 12.4 & 0.06 & 0.185 &~ 403 &~ 46  & 55 &70  &69\\
                &      & 0.12 & 0.234 &~ 252 &~ 41  & 60 &76  &75\\
                & 12.6 & 0.07 & 0.192 &~ 371 &~ 48  & 57 &73  &72\\
                &      & 0.14 & 0.239 &~ 238 &~ 44  & 65 &79  &77\\
\hline
H\cite{Radtke}  & 15.4 & 0.1  & 0.175 &~ 364 &~ 62  & 77  &95 & 94\\
                & 18.7 & 0.15 & 0.165 &~ 337 &~ 91  & 103 &125& 124\\
                & 21.5 & 0.25 & 0.170 &~ 276 &~ 114 & 131 &156& 155\\
\hline
\end{tabular}
\end{center}
\end{table}

We compare the results of the new interpolation formula
(\ref{interneu}) with the former one given in \cite{RRR}, the
Ichimaru-Tanaka fit formula \cite{Ichimaru}, and experimental data for
strongly coupled plasmas \cite{Ivanov,Popovic,Radtke} in Table
\ref{table}.  Taking into account an experimental error of about
$30\%$ we find a good agreement between theory and experiment.
However, complete correspondence with the experiments can not be
anticipated because of deviations from the Coulomb potential for the
electron-ion interaction in rare gases and the occurrence of neutral
particles (partial ionization) not included in the present theoretical
approach, which is focussed to the fully ionized Coulomb system.  The
Debye-Onsager relaxation effect leads to a reduction of the electrical
conductivity in the order of 5$\%$ at $\Gamma \approx 1$.  A direct
comparison with simulation results would be highly desirable. However,
the standard approach through the current-current autocorrelation
function is still limited by the small number of simulated particles
and the corresponding statistical uncertainties.

The validity range of the new interpolation formula (\ref{interneu})
is restricted to a parameter range where the formation of bound states
($\Theta \ge 1$) can be neglected. Partial ionization plays a crucial role
in low-temperature plasmas and can lead to a minimum in the isotherms
of the electrical conductivity, see Ref.\ \cite{RRN} for the case of
hydrogen. Bound states (atoms) are also important when evaluating the
hopping conductivity for fluid hydrogen near to the nonmetal-to-metal
transition at megabar pressures \cite{RRKR}. At present, the treatment 
of the dc conductivity for plasmas where bound states may occur is
performed within the model of the partially ionized plasma, where the
composition is determined within a thermodynamic approach, and the
contribution of the interactions between the different components is
considered separately. Within such an approach, the interpolation
formula given here may be of use to describe the contribution of the
interaction between electrons and ions to the conductivity.

The generalization of the present approach to plasmas with higher
charged ions $Z\ge1$ is also possible but not intended here. Then,
comparison with new experimental data for strongly coupled metal
plasmas with $\Gamma\gg1$ \cite{Benage,Kunze} can be performed, see
\cite{RK}. The inclusion of bound state formation and the extension to
higher charged ions should be considered as possible goals of future
work on interpolation formula for the electrical conductivity of
nonideal plasmas.

\begin{acknowledgements}
  This work was supported by the Deutsche Forschungsgemeinschaft
  within the Sonderforschungsbereich 198 {\it Kinetics of Partially
    Ionized Plasmas}. One of us (A.E.) thanks the University of
  Rostock for the kind hospitality.
\end{acknowledgements}


\begin{received}
\end{received}

\end{document}